\documentclass{article}
\usepackage[preprint]{spconf}
\usepackage{spconf,amsmath,graphicx, bbold}
\usepackage{hyperref}
\usepackage{url}

\title{Using IPA-Based Tacotron for Data Efficient Cross-Lingual Speaker Adaptation and Pronunciation Enhancement}
\name{Hamed Hemati, Damian Borth}
\address{University of St. Gallen}
\begin{document}
%
\maketitle
\begin{abstract}
Recent neural Text-to-Speech (TTS) models have been shown to perform very well when enough data is available. However, fine-tuning them for new speakers or languages is not straightforward in a low-resource setup. In this paper, we show that by applying minor modifications to a Tacotron model, one can transfer an existing TTS model for new speakers from the same or a different language using only 20 minutes of data. For this purpose, we first introduce a base multi-lingual Tacotron with language-agnostic input, then demonstrate how transfer learning is done for different scenarios of speaker adaptation without exploiting any pre-trained speaker encoder or code-switching technique. We evaluate the transferred model in both subjective and objective ways.

\end{abstract}


%
\section{Introduction}
\label{sec:intro}
\vspace{-0.2cm}
Recent TTS models \cite{shen2018natural, ren2020fastspeech, ping2018deep}  can generate intelligible and high fidelity speech when trained on tens of hours of speech corpus. However, training them with only a small amount of data results in an unstable TTS model. Different techniques have been proposed for pre-training parts of a TTS model as a pre-training step and then fine-tuning it towards new speakers \cite{chung2018semisupervised}. In such a case, however, the distinction between the pre-training and the main training phases of the model makes overall convergence more difficult.

The goal of one type of transfer learning in TTS systems called speaker adaption is to adapt an existing model to a new speaker with a limited amount of data. A common way to do speaker adaption is by using a separate speaker encoder and conditioning the speech synthesis on the speaker embedding generated by the speaker encoder. This allows the model to infer the speaker identity from the speaker embedding and makes it possible to extend speech synthesis to unseen speakers during inference \cite{jia2019transfer, zhang2019learning}. One drawback of such models is their dependency on large multi-speaker datasets for training the base TTS model. Further, it requires a comprehensive speaker embedder to learn as many voice characteristics as possible during training. 

Another way of speaker adaptation is based on fine-tuning the weights of an already trained model \cite{moss2020boffin, chen2019sample}. Assuming the text encoder part of the TTS model has learned ``good enough'' representations, one usually freezes parts of the model that are supposed to stay fixed and only fine-tunes the remaining parts corresponding to the speaker identity to adapt to the new speaker. As the focus in speaker adaptation has been mainly on monolingual transfer, usually, the text encoder is specific to one particular language and therefore can not be used for transfer learning to a new language. In \cite{liu2019crosslingual} the authors learn a mapping between the source and target language phoneme, which helps with language transfer; however, one needs a pre-trained ASR model to learn this mapping in an unsupervised manner.

Thanks to large, publicly available datasets, training a single-speaking TTS model in a single language like English can be seen as a straightforward task. However, the open question is if it would be possible to adapt a TTS model to a new speaker from a different language without needing an additional ASR model or speaker embedder. To answer this question, one needs to design the model such that it can accept inputs from the source and target languages at the same time without additional input modules or phoneme converters. 

In this work, inspired by the idea of "learning without forgetting" \cite{li2017learning}, we look at speaker adaptation from a model weight fine-tuning perspective by preserving the previous speaker(s) in several scenarios. We first introduce a framework based on Tacotron 2 \cite{shen2018natural} and enhance it with minor adaptations, which make the convergence of the model faster and more stable, especially for cross-lingual cases. The model receives IPA \footnote{IPA:https://www.internationalphoneticassociation.org} characters as direct input passed through a trainable lookup table for almost all IPA characters, which allows us to easily extend to a new language without using a different encoder per language or any language switching technique. 
Moreover, we investigate whether fine-tuning a model for a new speaker in a different language with preserving the old speaker(s) characteristics would help the old speaker(s) to speak in the new language intelligibly. Simply put, the main advantages of such a TTS system are:
\vspace{-0.3cm}
\label{sec:typestyle}
\begin{itemize}
\setlength\itemsep{0.20em}
    \item Adapting to new speakers in a low-resource setup (from the same or a different language).
    \item Enabling the source speaker to speak the target language intelligibly and vice-versa.
\end{itemize}

\vspace{-0.5cm}
\section{Framework}
In this section, we introduce the TTS model that we’ve used in the experiments throughout this paper.

\subsection{Model Architecture}
\label{sssec:subsubhead}

The framework is based on is an attention-based sequence-to-sequence model based on Tacotron's architecture. The main modules of the model are an encoder (text encoder) and decoder with an attention layer in between, a speaker embedding lookup table, and a phoneme embedding lookup table. The only change that we made to the original Tacotron model is adding a residual connection from the input layer to the encoder's outputs, similar to \cite{9054722}. We found that adding this residual connection helps with the robustness of the model during training and inference and causes faster convergence in the monolingual transfer learning scenario where the encoder is frozen. 

\begin{figure}
    \centering
    \begin{minipage}[b]{1.0\linewidth}
  \centering
  \centerline{\includegraphics[width=8.5cm]{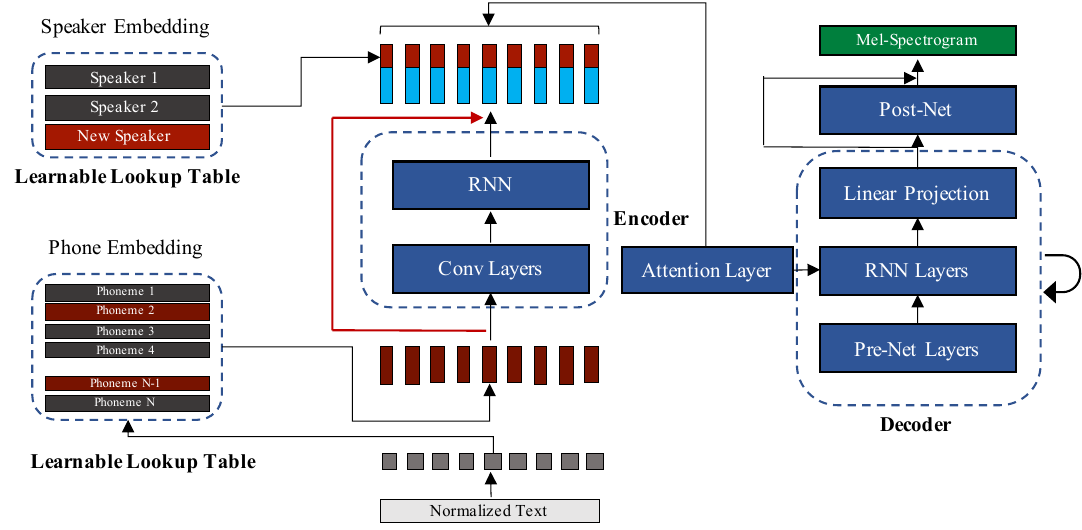}}
\end{minipage}
\caption{Architecture of the model. New speaker(s) and phoneme embeddings (in cross-lingual transfer) are shown in red.}
\label{fig:architecture}
\end{figure}

The input text is first normalized and converted to IPA phonemes; then, using a lookup table, each IPA character is converted to an n-dimensional vector. The text encoder, which consists of a number of convolutional layers and a recurrent layer (in our case, a bi-directional LSTM) on top, converts the input characters to sequence text embedding vectors. The speaker embedding extracted from the speaker embedding lookup table is then concatenated with the encoder's outputs. The decoder attends over the encoder's output to generate the Mel-spectrogram frames of the synthesized speech. We use Forward attention \cite{Zhang_2018} in the attention layer of our model. In Figure \ref{fig:architecture} we show the overall architecture of the model. 

Since the phoneme lookup table is supposed to be universal, it uses a fixed-size table containing all IPA character embedding vectors. Each source speaker may use a different subset of the table tensors depending on the source language. This allows us to easily extend an existing model to a new speaker of a different language by transferring the learned IPA embedding of the source language to the target language while allowing the model mainly to focus on the new characters that were not present in the source language.

\subsubsection{Training and Inference}
\label{sssec:subsubhead}
In simple words, training a TTS model is done by maximizing $P(Y|X; W)$ where $X$ and $Y$ are the input transcript and mel-spectrogram of the synthesized speech respectively. In our TTS model, we can expand the term as $P(Y|X, S_{id}; [W_{e}, W_{d}, W_{p}, W_{s}])$ where $S_{id}$ is the id(s) of the source speaker(s) and $W_{e}$, $W_{d}$, $W_{p}$ and $W_{s}$ are the parameters of the encoder, decoder, phoneme lookup weights ad speaker lookup weights respectively. 

First, we train a model on a single or multi-speaker dataset for a data-rich language. Once the model is fully converged on the source language, we have optimized weights of the different parts of the model indicated by ${W^*}_{e}$, ${W^*}_{d}$, ${W^*}_{p}$, ${W^*}_{s}$. We will use these parameters to explain the transfer learning scenarios in the following sections. Similar to Tacotron 2, we train the model by minimizing the loss term below in both pretraining and transfer learning steps:

\begin{equation}
    \mathcal{L} = ||Y - Y_{post} ||^2 + ||Y - Y_{pre}||^2 
\end{equation}

where $Y_{pre}$ and $Y_{pre}$ are the outputs of the decoder's linear projection and Post-Net respectively.

\begin{figure}
    \centering
    \begin{minipage}[b]{1.0\linewidth}
  \centering
  \centerline{\includegraphics[width=8.0cm]{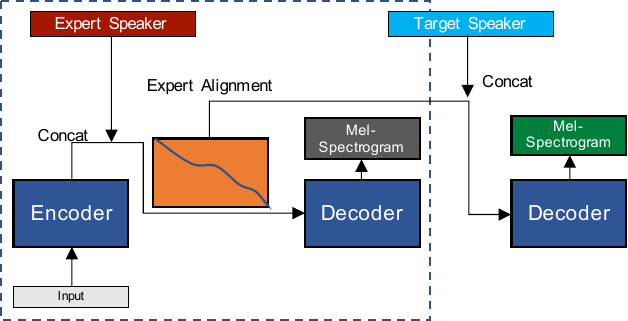}}
\end{minipage}
\caption{Expert alignment.}
\label{fig:expert_align}
\end{figure}

For cross-lingual speaker adaptation, we propose a 2-step inference mechanism which we call Expert Alignment. In expert alignment, as shown in Figure \ref{fig:expert_align}, first we compute the alignment with an (expert) speaker of the target language and then use the alignment for a speaker of a different language. We found this to be helpful for cross-lingual transfer with noisy datasets.
\label{sec:format}

\section{Transfer Learning Scenarios}
\label{sec:majhead}

A naive way of fine-tuning is to initialize the network weights for a new speaker with the optimal parameters obtained from training on a (single-speaker) large source dataset. However, once the number of sources and target speakers is different, we cannot just copy the weights from the source model. Moreover, even in the case of single-to-single speaker adaptation, simple fine-tuning will result in forgetting the source speaker, which doesn't allow us to improve the source speaker's pronunciation in the target language. Therefore, we keep both the origin speaker and add the target speaker(s) in our setup. In our experiments, we use 20 minutes of data per speaker. You can listen to demo examples from here\footnote{\href{https://d872c.github.io/ICASSP-2021_TTS}{\url{https://d872c.github.io/ICASSP-2021_TTS}}}. 

\subsection{Mono-Lingual Speaker Adaptation}
\label{sssec:subsubhead}

In monolingual speaker adaptation, we assume that the text encoder can already extract proper textual embedding from the input text. First, we froze different parts but noticed that both freezing and unfreezing the encoder and phoneme table resulted in almost the same performance. Nevertheless, since freezing keeps the encoder stable for noisy data, we kept the encoder frozen. In Figure 4(a), we show the Mel-cepstral distortion (MCD) of the training data in a monolingual transfer example for a model with and without encoder freezing. We did all monolingual transfer experiments with the objective below:
 \vspace{-0.20cm}
\begin{equation}
    \max P(Y|X, [\boldsymbol{S_{old}}, S_{new}]; [\boldsymbol{{W^*}_{e}}, {W^*}_{d}, \boldsymbol{{W^*}_{p}}, {W^*}_{s}])
\end{equation}

where the bold elements are frozen during transfer learning, and the rest of the parameters are updated. $S_{new}$ and $S_{old}$ correspond to the old and new speaker IDs, respectively.

\subsection{Cross-Lingual Speaker Adaptation}
\label{sssec:subsubhead}

In cross-lingual transfer learning, the target speaker is supposed to be from a different language than the source speaker. Since the target language might use a different combination of phonemes for word pronunciation, freezing the text encoder would cause pronunciation issues in the target language. For the speaker embeddings, similar to the mono-lingual case, we freeze the speaker embedding vector(s) for the old speaker(s) and only let the new speaker(s) adapt their embedding(s) to the TTS model. Therefore we use the objective below to do transfer learning for speaker(s) for a new language.

\begin{equation}
    \max  P(Y|X, [ S_{old}, S_{new}]; [{W^*}_{e}, {W^*}_{d}, {W^*}_{p}, \boldsymbol{{W^*}_{s}}])
\end{equation}
where the part written in bold is partially frozen.

\section{Experiment Setup}
\label{sec:print}
In this section, we explain the setup for different experiments that we did to evaluate the transferability of our TTS model. 

\subsection{Datasets}
\label{sssec:subsubhead}
We used single-speaker datasets to train our base model; however, one could also use multi-speaker datasets for this purpose. For the English source model, we used LJSpeech\cite{ljspeech17}, and for the German source model, we used our internal datasets, which contain one male and one female speaker that we will refer to as Inter-Male/Female. For the rest of the languages like Spanish, French, and Korean, we used CSS10\cite{park2019css10} which is publicly available.

For the transfer learning (speaker adaptation) part, we used two random speakers of VCTK and created a 20-min long noisy (imperfect-alignment) dataset from one of Donald Trump's speeches for the English language. For the German language, also we used 20-min long Anglea Merkel's voice.

\subsection{Baseline Model}
\label{sssec:subsubhead}

We used the publicly available source code for Tacotron 2 from NVidia \footnote{Tacotron 2 - NVidia: https://github.com/NVIDIA/DeepLearningExamples} as a baseline. We extended it with a residual connection as described in the Model Architecture section. For text-to-phoneme conversion, we used eSpeak NG \footnote{eSpeak NG: https://github.com/espeak-ng/espeak-ng} phonemizer which supports most languages. We trained each source model with batch-size 80 on a V100 GPU for almost ~60K iterations. 

\subsection{Evaluation Metrics}
\label{sssec:subsubhead}
We use both subjective and objective metrics for the evaluation of the transferred model. To evaluate the intelligibility of the generated sentences in terms of word pronunciation correctness fed the synthesized speech to a pre-trained ASR model based on DeepSpeech \footnote{DeepSpeech: https://gitlab.com/Jaco-Assistant/deepspeech-polyglot}. Then, we computed the WER and MER \cite{Morris2004FromWA} for the ASR model's output, the original file, and the generated speech from a model in one of the transfer learning models. We also computed the  MCD of the training data during the transfer learning as explained in \cite{fang2019transfer}.

The subjective evaluation was only done for the mono-lingual transfer case for English and German languages. We evaluated 20 different sentences for our three different speakers in both languages using the Amazon Mechanical Turk service. We used the interface and analysis tools provided by \cite{naderi2020} to acquire mean opinion scores (MOS) for our audio samples. A total of 117 workers participated in the evaluation, 48 of which were rejected due to misclassifications of our gold standard (trapping) clips, resulting in 20 workers for German 49 for English.

\section{Results}
\label{sec:print}

\subsection{Mono-Lingual Single Speaker Adaptation}
\label{sssec:subsubhead}

In mono-lingual adaptation, we transferred an English source model trained on LJSpeech to two speakers of VCTK and Trump's voice. The German source model trained on either InterMale or InterFemale was transferred as shown in Table 1. the model achieved reasonably good results after almost 3-4K iterations.

\begin{table}[]
\centering
\begin{tabular}{ |p{2.5cm}||p{0.7cm}|p{1.0cm}|p{0.7cm}|p{0.7cm}| }
 \hline

 \centering
  Mode  & Iters & MCD & WER & MOS\\
 \hline
 \hline
 LJ-to-VCTK298 & 3K  & $\approx$ 24.0 & 0.56 & 3.91 \\
 LJ-to-VCTK283 & 3K  & $\approx$ 23.0 & 0.59 & 3.55 \\
 LJ-to-Trump   & 3K  & $\approx$ 23.0 & 0.50 & 2.15 \\
 \hline
 Male-to-Female & 3K  & $\approx$ 22.0 & 0.16 & 3.85 \\
 Female-to-Male  & 3K & $\approx$ 25.0 & 0.14 & 3.19 \\
 Male-to-Merkel & 4K  & $\approx$ 21.0 & 0.25 & 3.68 \\
 \hline
\end{tabular}
\caption{\label{tab:table-name}Mono-Lingual Transfer}
\label{tab:monling}
\end{table}

\subsection{Cross-Lingual Speaker Adaptation}
\label{sssec:subsubhead}
For cross-lingual adaptation, we used LJSpeech and InterMale for the English and German source models, respectively, and the single speaker datasets from CSS10 for the rest of the languages. In Table 2, we show the results for adaptation from English, Korean, and Spanish to four different speakers in different languages. The scores are computed for the target speaker in the target language. 

We also tested the source speaker's pronunciation in the target language by sentences whose phonemization had low overlap with the target language. In Figure 4(b), as an example, we show how the attention for the Spanish speaker fails over an English sentence in the top row, and below, we show the attention after ES2EN speaker adaptation. 

\begin{table}
\centering
\begin{tabular}{ |p{1.5cm}||p{0.7cm}|p{1cm}|p{0.7cm}|p{0.7cm}| }
 \hline
 \multicolumn{5}{|c|}{Cross-lingual Speaker Adaptation} \\
 \hline
 \centering
  Mode    & Iters & MCD & WER & MER \\
 \hline
 EN2FR   & 7K & $\approx$ 23.0 & 0.58 & 0.49  \\
 EN2DE   & 8K & $\approx$ 20.5 & 0.12 & 0.12  \\
 EN2KO   & 6K & $\approx$ 24.0 & 0.70 & 0.64  \\
 EN2ES   & 5K & $\approx$ 21.5 & 0.10 & 0.10  \\
 \hline
 KO2FR   & 4K & $\approx$ 25.0 & 0.55 & 0.52  \\
 KO2EN   & 3K & $\approx$ 28.0 & 0.78 & 0.77 \\
 KO2DE   & 19K & $\approx$ 21.0 & 0.24 & 0.23  \\
 KO2ES   & 3K & $\approx$ 25.5 & 0.19 & 0.18  \\
 \hline
 ES2FR   & 4K & $\approx$ 23.0 & 0.50 & 0.44  \\
 ES2EN   & 10K & $\approx$ 20.0 & 0.25 & 0.23 \\
 ES2DE   & 6K & $\approx$ 21.5 & 0.19 & 0.15 \\
 ES2KO   & 14K & $\approx$ 21.0 & 0.88 & 0.78 \\
 \hline
\end{tabular}
\caption{\label{tab:table-name}Cross-Lingual Speaker Adaptation.}
\end{table}

\subsection{Cumulative Cross-Lingual Speaker Adaptation}
\label{sssec:subsubhead}
Another interesting scenario to consider is Cumulative language learning. The goal is to check whether the TTS model can be trained in a cumulative way of cross-lingual learning. For example, given a source model originally trained on English then adapted to a Spanish speaker, would it be able to learn a third and fourth language?

In Table 3, we show two random cumulative experiments. In general, the models converge faster than the bi-lingual case, probably due to the additional learned phoneme combinations and speech variations compared to the initial source model. In Figure 3, we visualize (with t-SNE) how the collapsed embeddings of the German IPA characters (in green) in an ES2KO model are expanded after adapting it to a German speaker. 

\begin{figure}
    \centering
    \begin{minipage}[b]{.48\linewidth}
  \centering
  \centerline{\includegraphics[width=4.0cm]{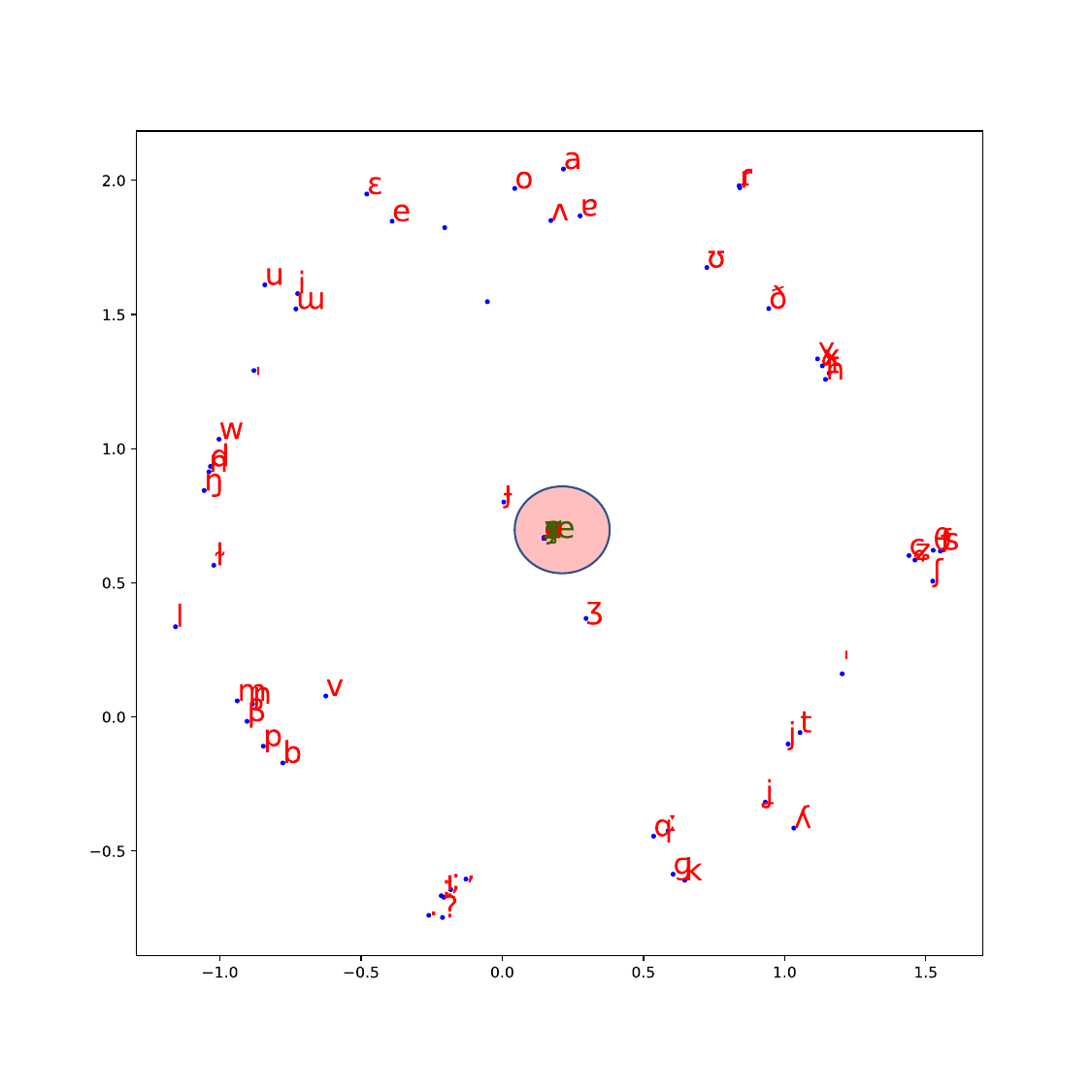}}
 \vspace{.1cm}
\end{minipage}
\hfill
\begin{minipage}[b]{0.48\linewidth}
  \centering
  \centerline{\includegraphics[width=4.0cm]{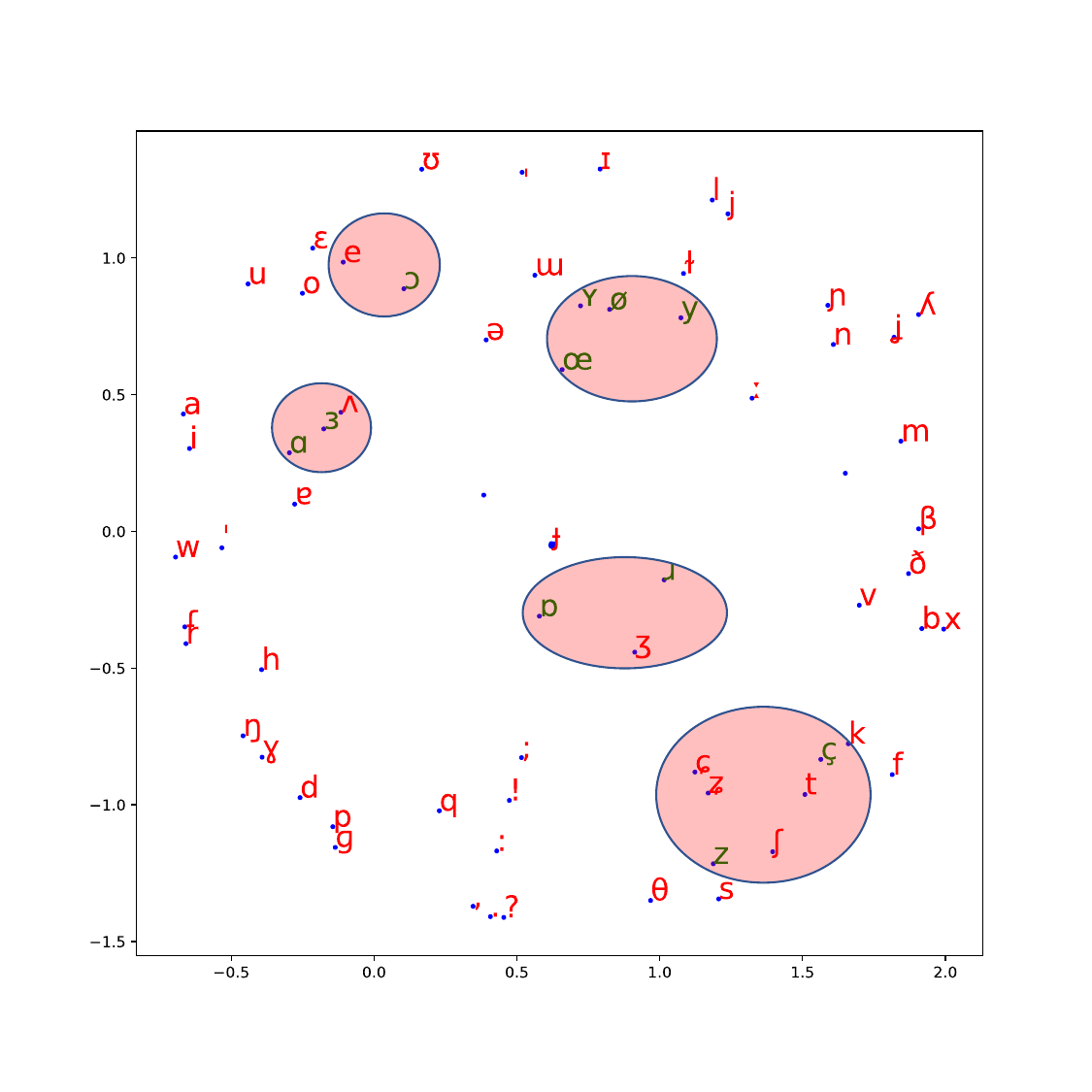}}
 \vspace{.1cm}
\end{minipage}
 \vspace{-.6cm}



 \vspace{-.2cm}

\caption{ES2KO $\rightarrow$ ES2KO2DE Speaker Adaptation}
\label{fig:res}
\end{figure}

\begin{table}
\centering
\begin{tabular}{ |p{1.7cm}||p{0.5cm}|p{1.0cm}|p{1.14cm}|p{1.14cm}| }
 \hline
 \centering
  Mode    & Iters & MCD & WER & MER \\
 \hline
 DE2ES2FR   & 3K & $\approx$ 24.5 & 0.61/0.71 & 0.55/0.69  \\
 ES2KO2DE   & 7K & $\approx$ 24.0 & 0.19/0.44 & 0.19/0.39 \\

 \hline
 
\end{tabular}
\caption{\label{tab:table-name}Cumulative Speaker Adaptation.}
\end{table}

\begin{figure}[htb]



%
\begin{minipage}[b]{.48\linewidth}
  \centering
  \centerline{\includegraphics[width=4.0cm]{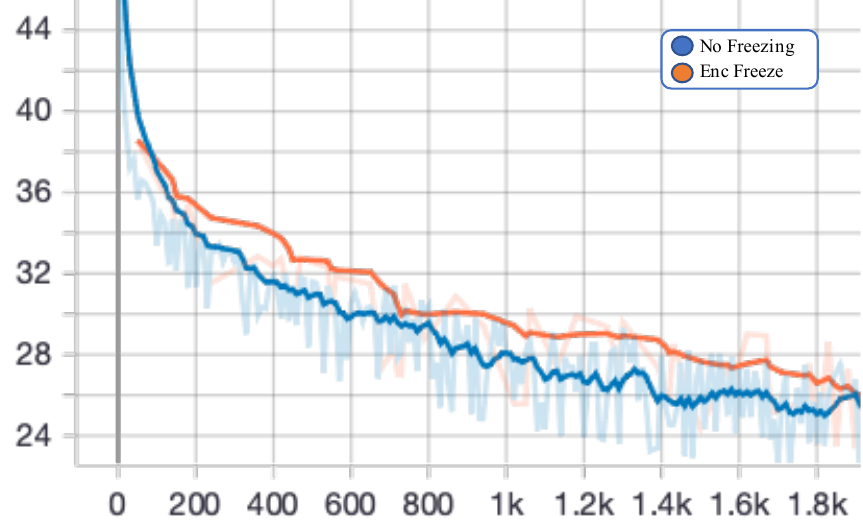}}
  \centerline{(a) No Freeze vs EncFreeze}\medskip
\end{minipage}
\hfill
\begin{minipage}[b]{0.48\linewidth}
  \centering
  \centerline{\includegraphics[width=4.0cm]{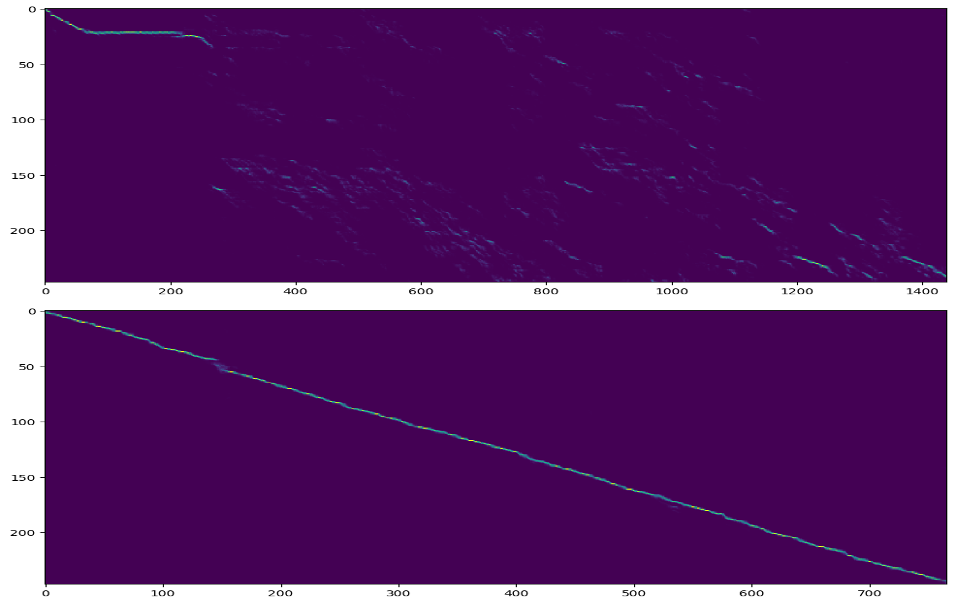}}
  \centerline{(b) ES Speaker's Attention}\medskip
\end{minipage}
\caption{Training and inference visualizations for monolingual and cross-lingual transfer learning setups.}
\label{fig:res}
\end{figure}

\section{Conclusion}
\label{sec:prior}
We propose simple modifications and fine-tuning techniques for Tacotron 2 in cross-lingual low-resource speaker adaptation setups. We show that by adopting a universal input definition and speaker embedding, it is possible to adapt to new speakers in a different language from a pre-trained model with a limited amount of data. We also conduct experiments and demonstrate that such adaptations result in better pronunciation of words in the target language for the source speaker and can be generalized to more languages in a cumulative manner.  

\vfill\pagebreak

\bibliographystyle{IEEEbib.bst}
\bibliography{strings}

\begin{thebibliography}{10}

\bibitem{shen2018natural}
Jonathan Shen, Ruoming Pang, Ron~J. Weiss, Mike Schuster, Navdeep Jaitly,
  Zongheng Yang, Zhifeng Chen, Yu~Zhang, Yuxuan Wang, RJ~Skerry-Ryan, Rif~A.
  Saurous, Yannis Agiomyrgiannakis, and Yonghui Wu,
\newblock ``Natural tts synthesis by conditioning wavenet on mel spectrogram
  predictions,'' 2018.

\bibitem{ren2020fastspeech}
Yi~Ren, Chenxu Hu, Xu~Tan, Tao Qin, Sheng Zhao, Zhou Zhao, and Tie-Yan Liu,
\newblock ``Fastspeech 2: Fast and high-quality end-to-end text to speech,''
  2020.

\bibitem{ping2018deep}
Wei Ping, Kainan Peng, Andrew Gibiansky, Sercan~O. Arik, Ajay Kannan, Sharan
  Narang, Jonathan Raiman, and John Miller,
\newblock ``Deep voice 3: Scaling text-to-speech with convolutional sequence
  learning,'' 2018.

\bibitem{chung2018semisupervised}
Yu-An Chung, Yuxuan Wang, Wei-Ning Hsu, Yu~Zhang, and RJ~Skerry-Ryan,
\newblock ``Semi-supervised training for improving data efficiency in
  end-to-end speech synthesis,'' 2018.

\bibitem{jia2019transfer}
Ye~Jia, Yu~Zhang, Ron~J. Weiss, Quan Wang, Jonathan Shen, Fei Ren, Zhifeng
  Chen, Patrick Nguyen, Ruoming Pang, Ignacio~Lopez Moreno, and Yonghui Wu,
\newblock ``Transfer learning from speaker verification to multispeaker
  text-to-speech synthesis,'' 2019.

\bibitem{zhang2019learning}
Yu~Zhang, Ron~J. Weiss, Heiga Zen, Yonghui Wu, Zhifeng Chen, RJ~Skerry-Ryan,
  Ye~Jia, Andrew Rosenberg, and Bhuvana Ramabhadran,
\newblock ``Learning to speak fluently in a foreign language: Multilingual
  speech synthesis and cross-language voice cloning,'' 2019.

\bibitem{moss2020boffin}
Henry~B. Moss, Vatsal Aggarwal, Nishant Prateek, Javier González, and Roberto
  Barra-Chicote,
\newblock ``Boffin tts: Few-shot speaker adaptation by bayesian optimization,''
  2020.

\bibitem{chen2019sample}
Yutian Chen, Yannis Assael, Brendan Shillingford, David Budden, Scott Reed,
  Heiga Zen, Quan Wang, Luis~C. Cobo, Andrew Trask, Ben Laurie, Caglar
  Gulcehre, Aäron van~den Oord, Oriol Vinyals, and Nando de~Freitas,
\newblock ``Sample efficient adaptive text-to-speech,'' 2019.

\bibitem{liu2019crosslingual}
Zhaoyu Liu and Brian Mak,
\newblock ``Cross-lingual multi-speaker text-to-speech synthesis for voice
  cloning without using parallel corpus for unseen speakers,'' 2019.

\bibitem{li2017learning}
Zhizhong Li and Derek Hoiem,
\newblock ``Learning without forgetting,'' 2017.

\bibitem{9054722}
X.~{Zhou}, X.~{Tian}, G.~{Lee}, R.~K. {Das}, and H.~{Li},
\newblock ``End-to-end code-switching tts with cross-lingual language model,''
\newblock in {\em ICASSP 2020 - 2020 IEEE International Conference on
  Acoustics, Speech and Signal Processing (ICASSP)}, 2020, pp. 7614--7618.

\bibitem{Zhang_2018}
Jing-Xuan Zhang, Zhen-Hua Ling, and Li-Rong Dai,
\newblock ``Forward attention in sequence- to-sequence acoustic modeling for
  speech synthesis,''
\newblock {\em 2018 IEEE International Conference on Acoustics, Speech and
  Signal Processing (ICASSP)}, Apr 2018.

\bibitem{ljspeech17}
Keith Ito and Linda Johnson,
\newblock ``The lj speech dataset,''
  \url{https://keithito.com/LJ-Speech-Dataset/}, 2017.

\bibitem{park2019css10}
Kyubyong Park and Thomas Mulc,
\newblock ``Css10: A collection of single speaker speech datasets for 10
  languages,''
\newblock {\em Interspeech}, 2019.

\bibitem{Morris2004FromWA}
A.~Morris, V.~Maier, and P.~Green,
\newblock ``From wer and ril to mer and wil: improved evaluation measures for
  connected speech recognition,''
\newblock in {\em INTERSPEECH}, 2004.

\bibitem{fang2019transfer}
Wei Fang, Yu-An Chung, and James Glass,
\newblock ``Towards transfer learning for end-to-end speech synthesis from deep
  pre-trained language models,'' 2019.

\bibitem{naderi2020}
Babak Naderi and Ross Cutler,
\newblock ``An open source implementation of itu-t recommendation p. 808 with
  validation,''
\newblock {\em Proc. Interspeech}, 2020.

\end{thebibliography}

\end{document}